\documentclass{kluwer}    % Specifies the document style.

\usepackage{epsfig}

\newdisplay{guess}{Conjecture}

\newcommand{\citeP}[1]{(\citeauthor{#1} \cite*{#1})}
\newcommand{\citeN}[1]{\citeauthor{#1} (\cite*{#1})}
\newcommand{\citeNP}[1]{\citeauthor{#1} \cite*{#1}}

\newcommand{\apj} {ApJ}
\newcommand{\apjs} {ApJS}
\newcommand{\apjl} {ApJL}
\newcommand{\aap} {A\&A}
\newcommand{\solphys} {Solar Physics}
\newcommand{\arcsec} {''}

\begin{document}

\begin{article}
\begin{opening}         
\title{Spectro-Polarimetric Observations and Non-LTE Modeling of Ellerman
  Bombs}  
\author{Hector \surname{Socas-Navarro}, Valent\' \i n {Mart\' \i nez Pillet},
  David \surname{Elmore}, Anna 
  \surname{Pietarila}, Bruce W. \surname{Lites} \& Rafael \surname{Manso
    Sainz} \thanks{Visiting
  Astronomers, National Solar Observatory, 
    operated by the Association of Universities for Research in Astronomy,
    Inc. (AURA), under cooperative agreement with the National Science
    Foundation.}}
\runningauthor{Socas-Navarro et al}
\runningtitle{Spectro-Polarimetric Observations of Ellerman Bombs}
\institute{High Altitude Observatory, NCAR \thanks{The National Center for
  Atmospheric Research (NCAR) is sponsored by the National Science
  Foundation}} 
\date{Feb 16, 2005}

\begin{abstract}
Ellerman bombs are bright emission features observed in the wings of
H$\alpha$, usually in the vicinity of magnetic concentrations. Here we show
that they can also be detected in the Ca~II infrared triplet lines, which are
easier to interpret and therefore allow for more detailed diagnostics. We
present full Stokes observations of the 849.8 and 854.2~nm lines acquired
with the new spectro-polarimeter SPINOR. The data shows no significant linear
polarization at the level of 3$\times$10$^{-4}$. The circular polarization
profiles exhibit measureable signals with a very intricate pattern of
peaks. A non-LTE analysis of the spectral profiles emerging from these
features reveal the presence of strong downflows ($\sim$10~km/s) in a hot
layer between the upper photosphere and the lower chromosphere.
\end{abstract}

\end{opening}

\section{Introduction}
\label{sec:intro}

One of the most intriguing discoveries made in Solar Physics early last century 
was the phenomenon known as Ellerman bombs. 
The common use of the H$\alpha$ line to monitor the Sun
allowed the discovery of features with enhanced emission peaks 
in the wings of this line, not affecting the central
absorption core (\citeNP{E17}). Off-band ($\sim$ 1 \AA) H$\alpha$ filtergrams
showed these features to be spatially point-like structures with a size of
$\sim$1\arcsec, lasting for little more than 10 minutes (see \citeNP{Q00}
and references therein). As they are virtually absent from line core
filtergrams, they have always been associated to events taking place in the
low chromosphere. This is also confirmed by their clear correlation with
brightenings seen in the 1600 \AA~continuum images from the TRACE satellite 
(\citeNP{Q00}). Sometimes, an X-ray transient brightening can also be 
observed at the same location (\citeNP{S02}). Recent reviews on Ellerman bombs 
are given by \citeN{R01} and in the introduction of \citeN{G02}.

The physical process behind Ellerman bombs has remained elusive. 
The blue asymmetry observed by the H$\alpha$ profiles (excess emission
in the blue wing) has sometimes been associated to outward
motions of the solar material \citeP{K70}. But using two-component models
separated along the line of sight, \citeN{D97} showed that the asymmetry could
be originated from, either, hot emitting material moving upwards
or cold absorbing material falling from the top of the chromosphere.
The relation to the magnetic field configuration has been
controversial as well. Ellerman bombs have been detected in two distinct
solar magnetic scenarios, but always inside active regions. They
occur in emerging regions between the opposite
polarity footpoints and also surrounding isolated sunspots, beyond the penumbral
boundary. For this later case, \citeN{N98} have shown that they can be
identified with the so-called Moving Magnetic Features (MMFs) that are seen 
in the moat region around sunspots (for a description of MMFs see \citeNP{H73}).
Whereas this points towards a magnetic origin of
Ellerman bombs, a number of events observed by \citeN{N98} (see also
\citeNP{D97}) were not clearly linked to magnetic structures. These
authors point out that a number of cases showed
a relationship with the boundaries of magnetic regions but others
did not.

More recently the situation has been clarified by the analysis of the
observations made by the Flare Genesis balloon experiment (\citeNP{G02}, 
\citeNP{P04}). Observing an emerging flux region, these authors describe
dipolar features that appear in the middle of the region, where the flux
emerges at the surface (below the so-called arch filament system). With
better sensitivity magnetograms than the previous works, 
\citeN{G02} found that almost all Ellerman
bombs studied could be identified with the neutral line above the dipolar
features (or in neutral lines inside supergranular boundaries). Those cases not 
linked with the presence of neutral lines could be associated with
interacting field lines above the photosphere, a result confirmed by the
extrapolations made by \citeN{P04}. These authors proposed, then, as the
most probably mechanism for the generation of Ellermam bombs some
form of magnetic reconnection in the low chromosphere. That reconnection 
happens efficiently at these heights is supported by the fact that the
temperature minimum is the region with the lowest electrical
conductivity, and resistive processes occur there more easily
than elsewhere (\citeNP{L99}). While the Flare Genesis observations correspond
to the central portion of an emerging active region, these 
conclusions offer an
interesting parallelism with the Ellerman bombs observed surrounding
sunspots and associated to MMFs. \citeN{P04} pointed out that many
of the bombs studied were spatially located at the dipped
portions of undulatory field lines in the flux emerging area. This
undulatory or serpentine configuration is also used for the explanation
of MMFs (as first proposed by \citeNP{H73}). In this way, a unifying scenario
for the generation of these point-like energy release processes could
be constructed. But a study similar to that made by \citeN{P04} for the
moat of sunspots is still lacking. The comparison between the Ellerman
bombs seen in these two scenarios and the determination of their physical
origin awaits for more solid observational tools than H$\alpha$ line
profiles.

In particular, techniques providing information of the magnetic
field at the chromospheric heights where the energy release
takes place, will be of great value. H$\alpha$ linear polarization 
signals observed in Ellerman bombs have been reported 
in the past (\citeNP{R92}), but their
exact amount an origin remains controversial. In this paper, we propose
the use of full Stokes spectropolarimetry in more reliable
diagnostic lines such as the CaII infrared triplet lines. The most
important advantage these lines offer is the availability of powerful
inversion techniques that can be used to cover almost the full chromospheric
region. This paper presents a first characterization of the
atmosphere of two regions
close to a sunspot that display the properties ascribed to the
Ellerman bombs.

\section{Observations}
\label{sec:obs}

The observations used in this work are first-light data from the new
instrument Spectro-Polarimeter for INfrared and Optical Regions (SPINOR, see
\citeNP{SNEP+05}), at the Dunn Solar Telescope (DST) located at the
Sacramento Peak 
observatory. The dataset contains the two chromospheric CaII lines at 849.8
and 854.2~nm and two photospheric FeI lines at 849.7 and 853.8~nm. We
observed active region NOAA 0634 on June 16 at 15:16 UT, carrying out a
350-step scan with a spatial stepping of 0.22\arcsec. By using the adaptive
optics system of the DST \cite{RHR+03}, we attained a spatial resolution as
good as 0.6\arcsec, although this figure varies slightly along the scanning
direction (the horizontal direction in the figures) due to temporal seeing
variations. 

In order to improve the signal-to-noise ratio in the Stokes profiles, we 
averaged the observations in 3$\times$3 pixel boxes. The noise is thus
reduced to $\sim$3$\times$10$^{-4}$ times the continuum intensity. At this
level there is barely no signal in Stokes~$Q$ and~$U$, so we decided to leave
them out of our analysis. The chromospheric Stokes~$V$ profiles, however,
reveal fascinatingly complex shapes with multiple peaks, while the
photospheric lines exhibit the usual antisymmetric profiles.

\begin{figure*}
\includegraphics[width=\maxfloatwidth]{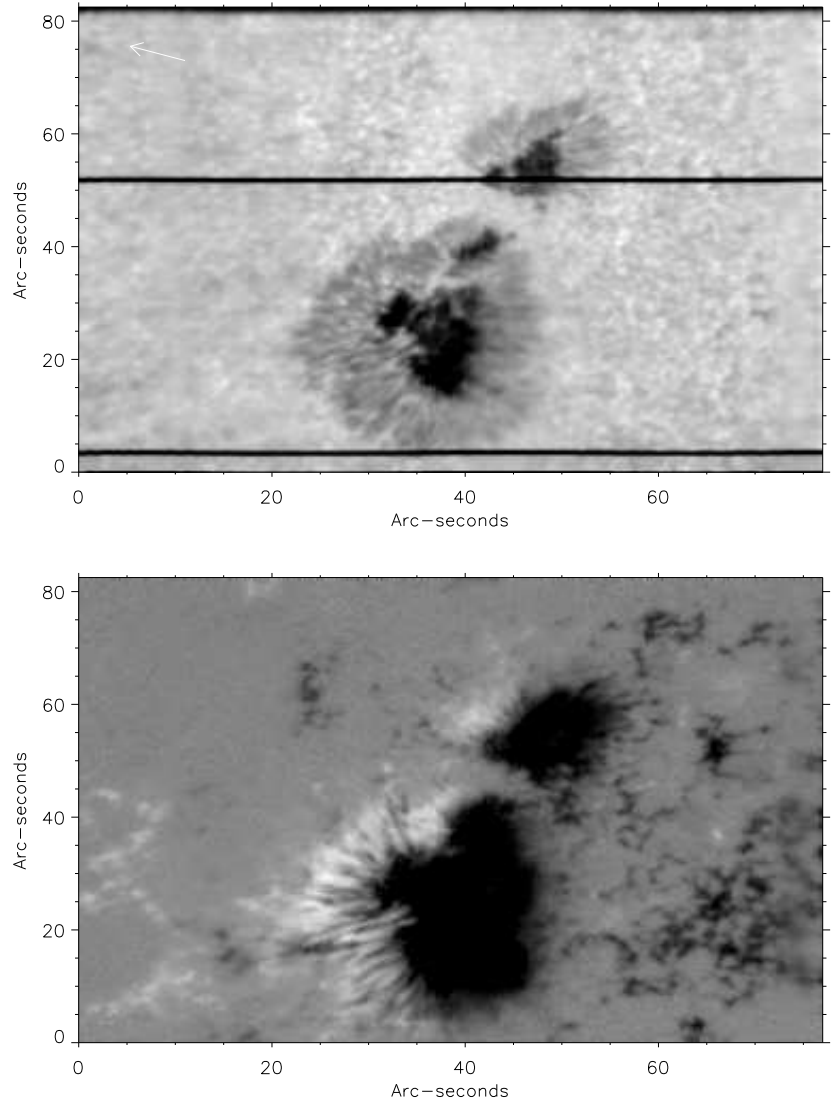}
\caption{Top: Continuum map reconstructed from the spectro-polarimetric
  observations. North is up and west is left of the image. The arrow
  indicates the direction to the solar limb. Bottom: Circular polarization in
  the red lobe of the photospheric Fe~I line at 8497 \AA .
  The horizontal dark stripes are hairlines inserted in the slit to serve as a
  reference for the calibration. 
}
\label{fig:maps}
\end{figure*}

\begin{figure*}
\includegraphics[width=1.3\maxfloatwidth]{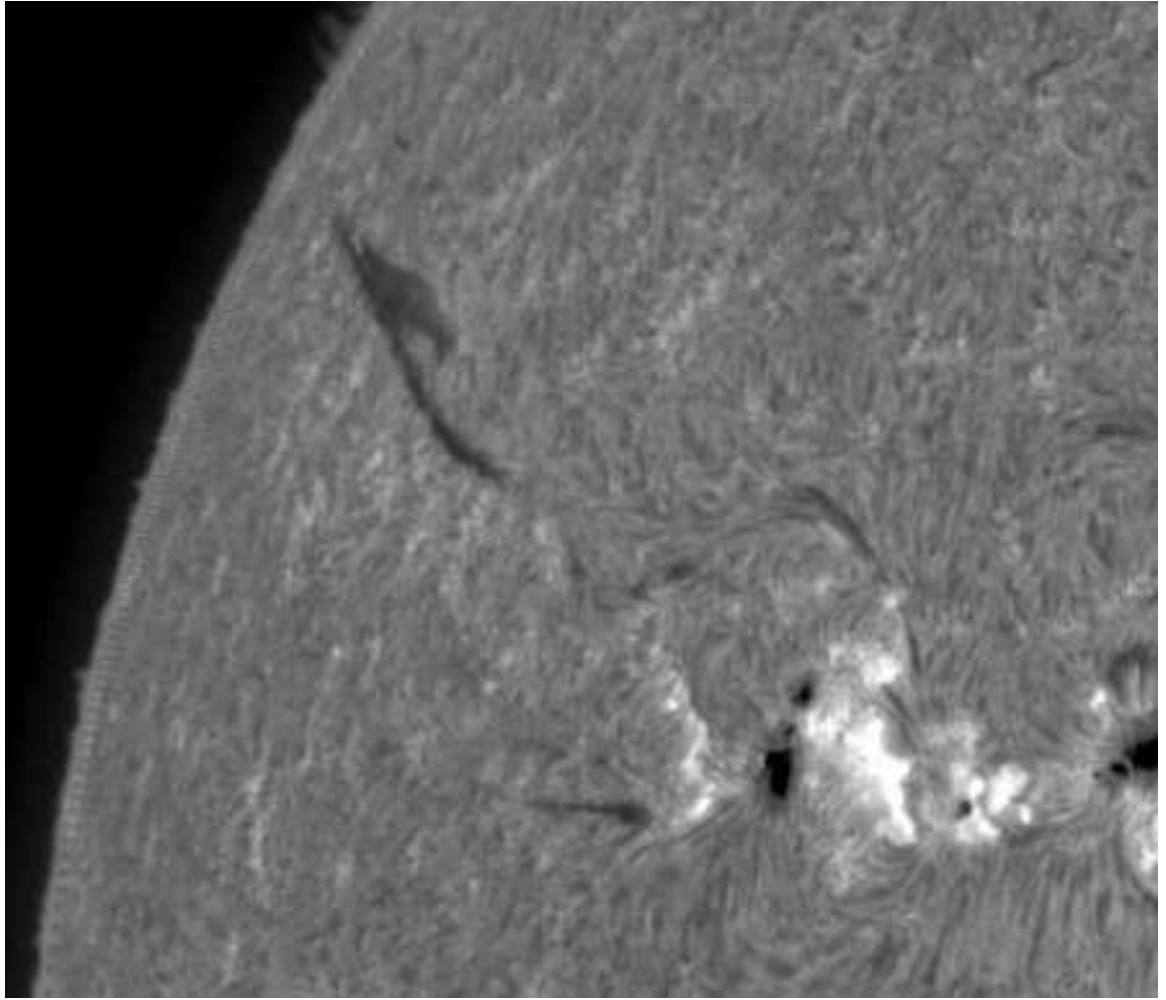}
\caption{BBSO H$\alpha$ image of NOAA 0634 taken on 16 June 2004 at 15:50
  UT. Notice the bright emission on the western side of the sunspot, near the
  penumbra.
}
\label{fig:Ha}
\end{figure*}

Fig~\ref{fig:maps} shows maps of the continuum intensity and a
``magnetogram'' taken in one of the photospheric lines. The magnetic
configuration surrounding the sunspots appears very different on both sides
of it. The eastern side exhibits a much more complex topology with an
intricate pattern of plage fields extending from the sunspot boundary to the
edge of the field of view. H$\alpha$ images of the region (Fig~\ref{fig:Ha})
reveal considerable emission near the eastern penumbral boundary. 

%IMPORTANT NOTE: The right (WEST) side of the penumbra is permeated by very
%strong currents in the photosphere (\citeNP{SN05d}).
%Since this is the area that exhibits the Ellerman bombs
%(described below), the bombs might be caused by magnetic reconnection on this
%side. Notice that the magnetic topology looks more complex on this side than
%on the left side of the figure, supporting the relation between bombs and
%reconnection. Also, the H$\alpha$ images (below) show a lot of activity on
%the right side of the figure. Perhaps this discussion should be in the
%conclusions but it might be worth to advance something here, so that the
%reader notices the difference in the E and W side of the spot as evidenced in
%this magnetogram.

\subsection{Velocities}

Photospheric velocity maps of NOAA~0634 were obtained by measuring the
minimum position of the Fe~I line at 849.7~nm. A similar strategy is
not viable for the Ca~II lines, however, because of the complicated
patterns of emission reversals and self-absortions found in the line
cores. Instead, we measured the intensity difference between two points
symmetrically located at various distances from the line core. This
method is only an approximation but it works better than
taking the minimum position.

\begin{figure*}
\includegraphics[width=\maxfloatwidth]{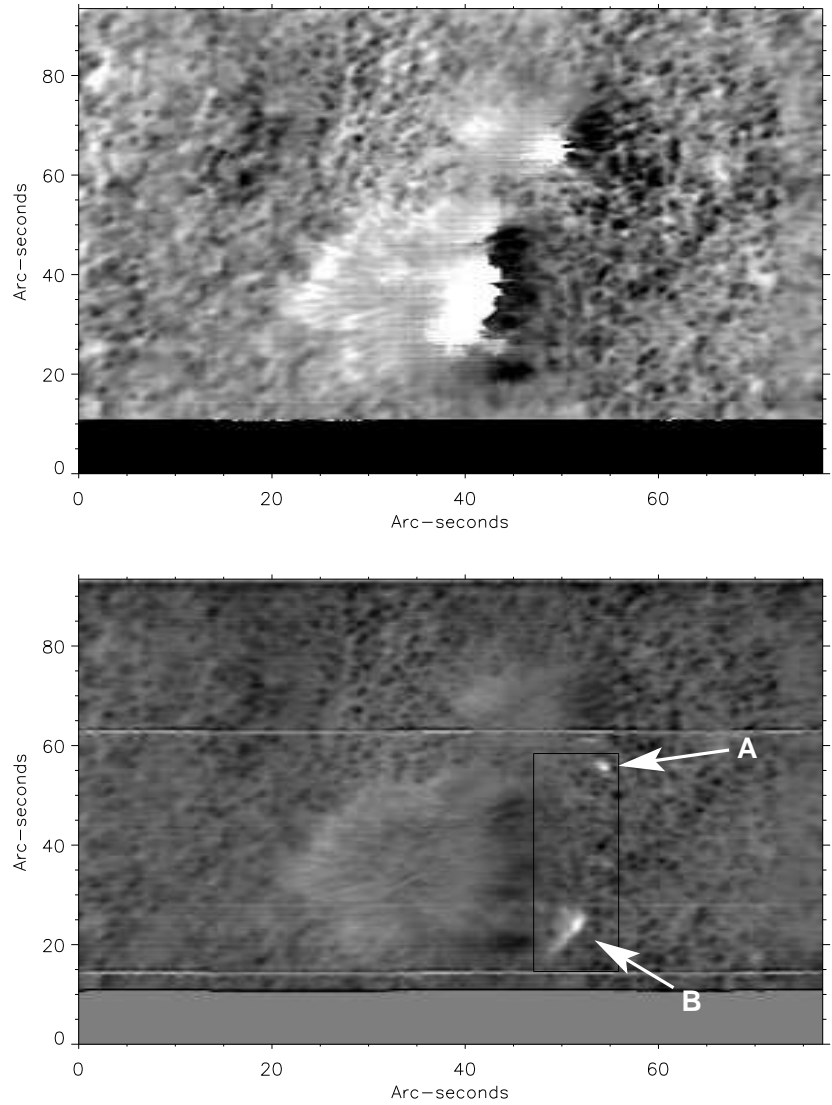}
\caption{
Top: Velocity map (arbitrary units) measured as the shift of Fe~I
 $\lambda\lambda$8497 line core. In this paper we follow the Astrophysical
 convention that  positive values correspond to redshifts. Bottom: 
 Dopplergram obtained by subtracting the intensity at two symmetric
  positions of the Ca~II line at 8498 \AA . The positions measured are 1.35
 \AA  \, apart, on the wings of the Ca~II line. The photospheric character of
 both maps is evident from the granulation pattern. The areas marked by
 arrows and labeled ``A'' and ``B'' in 
 the figure exhibit strong ($\sim$10~km~s$^{-1}$) flows and have been subject
 to detailed analysis in this work. 
}
\label{fig:vels}
\end{figure*}

\begin{figure*}
\includegraphics[width=\maxfloatwidth]{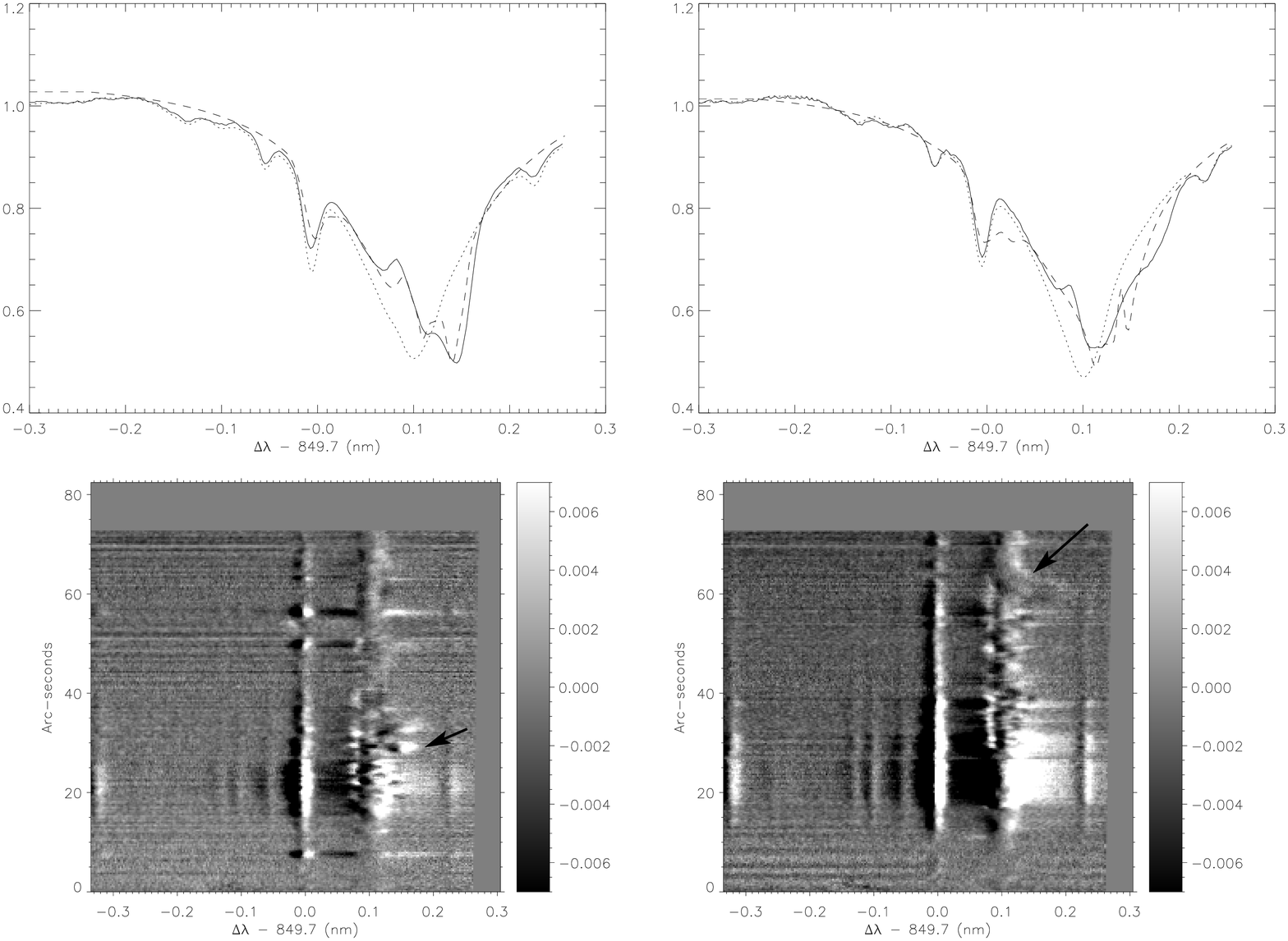}
\caption{Top: Two sample profiles of the Ca~II line at 8498~\AA \,
  observed in areas A (left) and B (right). These profiles exhibit
  particularly striking downflows. The dashed line shows the fit from the
  non-LTE inversion. Overplotted in dotted line for reference is the
  average profile along the slit direction. Bottom: Stokes~V spectra along
  the slit. The arrows mark the position of the profiles shown in the upper 
  panels. 
}
\label{fig:velprofs}
\end{figure*}

Fig~\ref{fig:vels} shows the velocity maps obtained at two heights, with the
convention that positive velocities are away from the observer. The
cosine of the heliocentric angle for this region is $\mu=0.71$. Notice the
two prominent bright features marked with arrows in the lower panel of the
figure. These features are strong plasma flows directed away from the
line of sight and located on the center side of the sunspot (but outside
the photospheric penumbral boundary). Detailed profiles of these features are 
shown in Fig~\ref{fig:velprofs}.

Both features exhibit very strong redshifts, visible in Stokes~$I$
and~$V$. Point~A shows a very prominent redshift starting near 40~pm
away from line center (corresponding roughly to the high photosphere
or low chromosphere) all the way to the line core (middle
chromosphere). This feature is well localized (see lower panel) and
the Doppler shift is $\sim$14~km~s$^{-1}$. Point~B exhibits somewhat
weaker flows ($\sim$10~km~s$^{-1}$) that are more localized in
height (upper photosphere/lower chromosphere), but has a more extended
tail along the slit (see lower panel). Notice how both points~A and~B
coincide with H$\alpha$ emissions in Fig~\ref{fig:Ha}

%The polarization signal indicates that these features are of magnetic
%nature, perhaps the result of chromospheric plasma falling down at
%high speed inside a magnetic filament. Further observations with time
%resolution and detailed Non-LTE modeling would be desirable to
%understand the structure and evolution of such strong downflows.

\section{Non-LTE inversions}
\label{sec:inv}

We performed detailed non-LTE inversions of the area surrounding points~A and~B
%inside the rectangle highlighted 
in Fig~\ref{fig:vels}. The inversions were
carried out with the code of \citeN{SNTBRC00a} (see also \citeNP{SNRCTB98})
by applying it systematically on each pixel independently. In order to
minimize the risk of the algorithm settling on a secondary minimum, we
repeated each inversion 10 times with randomized initializations. The model
atmosphere that yields the best fit to the observed Stokes profiles is taken
as a good representation of the solar atmosphere under study. 
%A parallel
%scheme was implemented to run the code simultaneously on four different Linux
%workstations with a total of 8 Intel P4 processors (although not fully
%dedicated). On average we employed the equivalent of six 2~GHz processors for
%a total computing time of approximately 8 days.

The inversion code treats the CaII lines in non-LTE and the FeI lines in
LTE. Blends are treated consistently by adding the opacities of all the
relevant transitions at each wavelength and spatial point, and then solving
the radiative transfer equation with the total opacity. The non-LTE calculation
is based on the preconditioning strategy of \citeN{RH92} (see also
\citeNP{SNTB97}) assuming complete angle and frequency redistribution. This
last approximation is particularly good for the CaII infrared triplet lines,
as shown by \citeN{U89}. 

\begin{figure*}
\includegraphics[width=\maxfloatwidth]{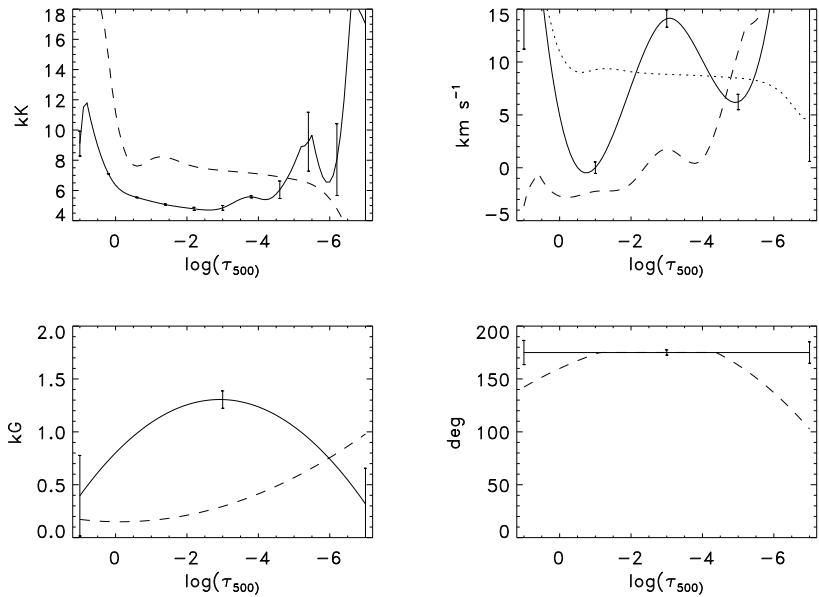}
\caption{Average model atmosphere in the downflowing feature labeled as
  point~A (solid) and its surroundings (dashed). Panels from top to bottom,
  left to right are temperature, line-of-sight velocity, magnetic field
  strength and inclination. The dotted line in the upper right panel
  represents the speed of sound in the atmosphere. Error bars are plotted at
  the {\it inversion nodes} (see Socas-Navarro et al 2000; 2001 for 
  details). 
}
\label{fig:modelsa}
\end{figure*}

\begin{figure*}
\includegraphics[width=\maxfloatwidth]{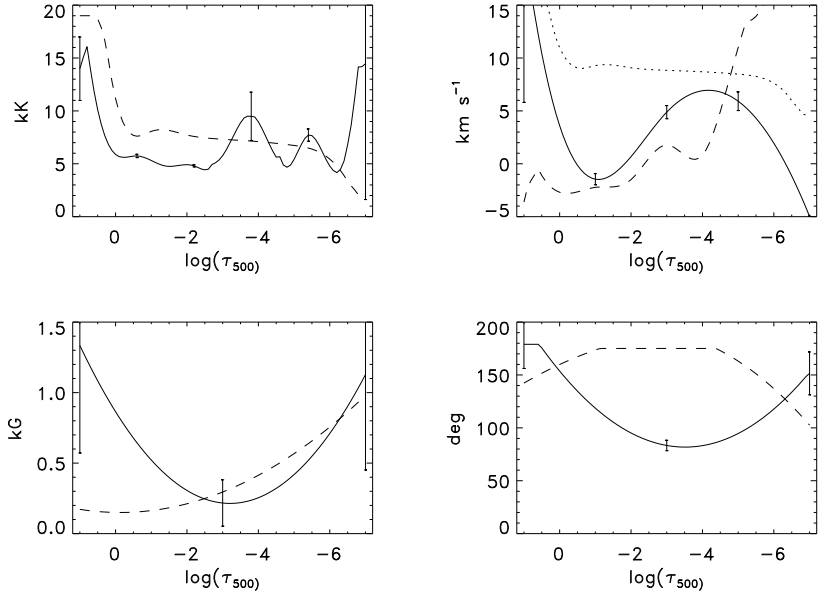}
\caption{Average model atmosphere in the downflowing feature labeled as
  point~B (solid) and its surroundings (dashed). Panels from top to bottom,
  left to right are temperature, line-of-sight velocity, magnetic field
  strength and inclination. The dotted line in the upper right panel
  represents the speed of sound in the atmosphere.
}
\label{fig:modelsb}
\end{figure*}

Let us consider in detail the models obtained for the points that we labeled
as~A and~B. Figures~\ref{fig:modelsa} and~\ref{fig:modelsb} show the
atmospheric stratification inside the downflows, as well as an average over
the surrounding, non-downflowing area for reference. The most striking
features in these models are the strong downwards velocities found in the
upper photosphere, around $\log(\tau_{500})=-3$. The 
motions along the line-of-sight approach the speed of sound in point~B and
exceed it in point~A. The other parameters show different behaviors in
points~A and~B. In particular, the downflowing layer in point~B is
considerably hotter than the reference atmosphere. However, in point~A the
downflowing layer is cooler (although it then becomes hotter at higher
layers). In both cases there is another hot layer at
$\log(\tau_{500})=-5$. Our findings for the two regions are consistent with
the results of \citeN{G02} who found Ellermam bombs to be predominantly
dominted by downflows with no indication of upflows.

\begin{figure*}
\includegraphics[width=\maxfloatwidth]{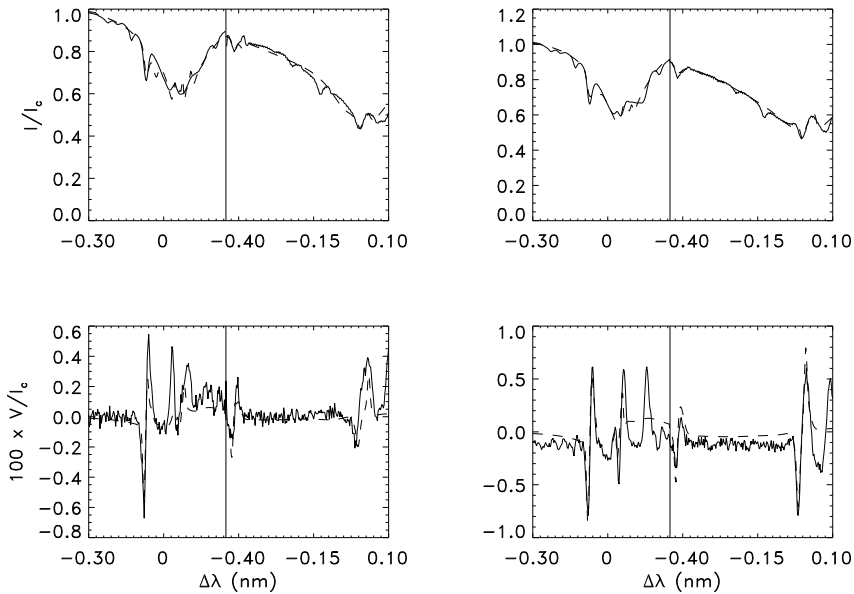}
\caption{Fits to 849.8 and 854.2~nm (left and right side of each panel,
  respectively). Left and right panels represent typical profiles in the
  downflowing areas in the vicinity of points~A and~B respectively. Upper
  panels: Stokes~I. Lower panels: Stokes~V. In all cases the solid (dashed)
  line represents the observations (fits). The profiles in this figure have
  been averaged over a 3$\times$3 pixel box to improve the signal-to-noise
  ratio in Stokes~$V$.
}
\label{fig:profs}
\end{figure*}

The fits provided by the inversions can be seen in Fig~\ref{fig:profs}. Note
that the observed chromospheric profiles exhibit a very complex structure
with multiple peaks in Stokes~$V$ (the photospheric lines, on the other hand,
produce rather normal profiles). Generally speaking, the inversion code
yields a reasonably good fit to the overall features, but there is more
structure in the observed profiles (particularly noticeable in Stokes~$V$
with multiple peaks), than in the synthetic ones. This is an indication that
the actual atmospheres are even more complex than our models, probably due to
the presence of unresolved atmospheric components.

\begin{figure*}
\includegraphics[width=\maxfloatwidth]{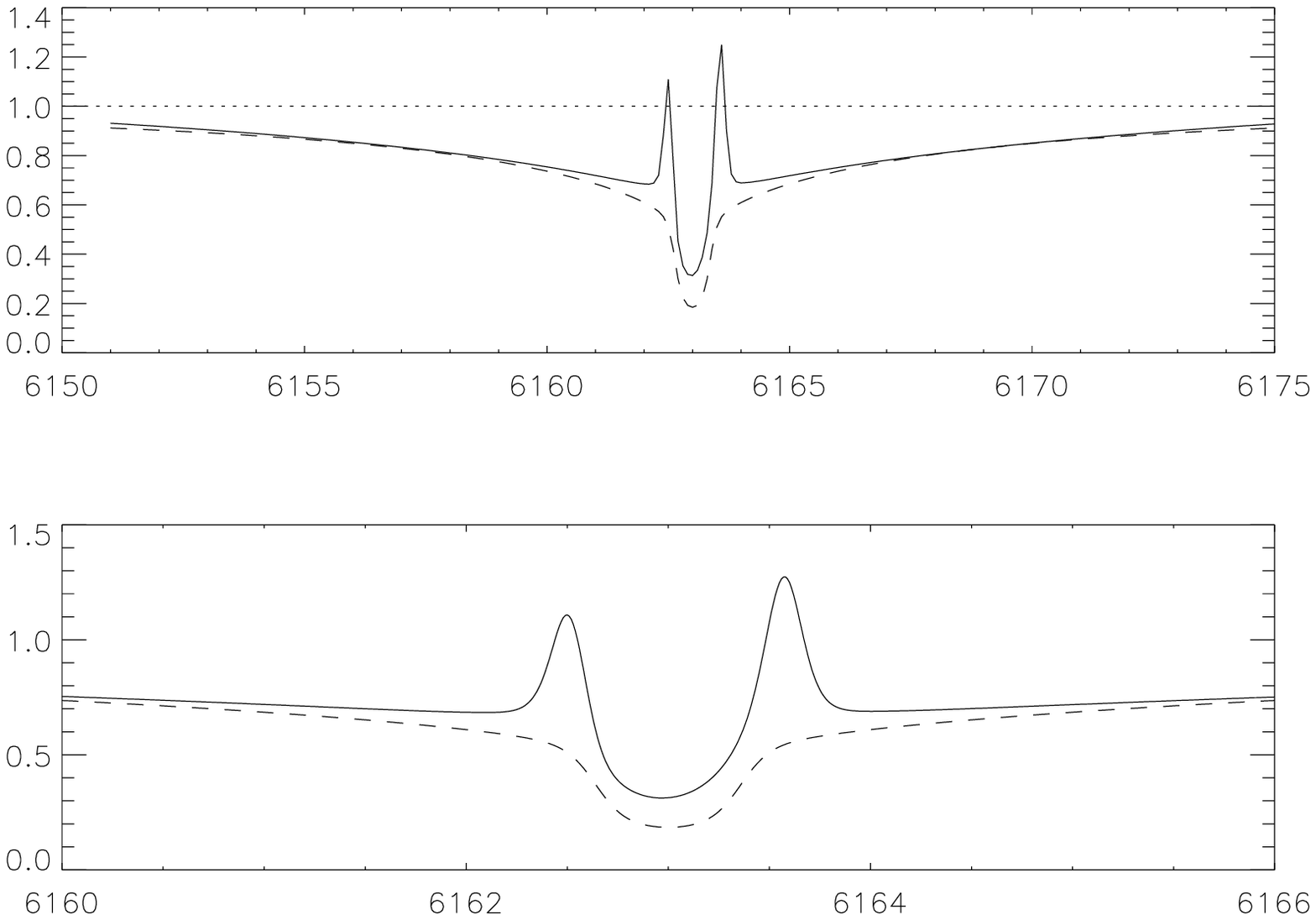}
\caption{Synthetic H$\alpha$ profiles obtained with the model atmosphere from
  the inversion of point~A (solid) and the VAL-C model (dashed). Point~B
  produced a qualitatively similar profile.
}
\label{fig:Halpha}
\end{figure*}

Thus far, we have not made a direct connection between the downflowing
features analyzed here and EBs. To this aim, we used the model atmospheres
obtained from the inversion to synthesize the H$\alpha$ profiles that these
models produce. The result is plotted in Fig~\ref{fig:Halpha}, along with a
reference quiet Sun profile calculated from model C of \citeN{VAL81}
(VAL-C). This synthetic profile has the same appearance traditionally
observed in EBs. \citeN{E17} describes the H$\alpha$ spectra of his bombs as
having emission features (brighter than the continuum) on both sides of the
line center. The red asymmetry (higher intensity peak in the red wing) shown
here is not the one most commonly observed in Ellerman bombs 
(see the introduction),
but this asymmetry is also present quite often (40\% of the cases according to
\citeNP{R01}). It is the result of a hot downflowing component as
seen in Fig \ref{fig:modelsa}. 
%Point B has a cold downflowing component and,
%thus, shows the more common blue asymmetry.
The resemblance of EBs with the synthetic profile in Fig~\ref{fig:Halpha} is
a solid evidence that we are indeed observing the same phenomenon. Moreover,
it also provides strong support for the reliability of the inversions.
%The spatial
%distribution of the physical properties is shown in Fig~\ref{fig:frames}. The
%figure displays the temperatures, magnetic field strengths and velocities
%retrieved at six different heights in the atmosphere.

%\begin{figure*}
%\includegraphics[width=1.3\maxfloatwidth]{frames.eps}
%\caption{Temperature (top panels), field strength (middle panels) and
%  line-of-sight velocity (bottom panels) obtained from the NLTE
%  inversions. Each panel displays the physical parameters in the rectangle
%  highlighted in Fig~\ref{fig:vels} above, although only a smaller subset
%  containing areas A and B has been inverted. Seven different heights in the
%  atmosphere are shown in this figure, increasing from left to right.  The
%  color scale is the same for all panels in each row to allow
%  for easy comparison of variables at different heights.
%\label{fig:frames}
%}
%\end{figure*}

\section{Conclusions}
\label{sec:conc}

The energetic phenomenon generally referred to as EBs has remained very
ellusive to solar physicists for almost a century. This is probably due in no
small measure to the fact that it has only been observed in H$\alpha$ thus
far, which is a very difficult line to interpret. In this paper we present
the first observations of EBs in other lines, namely the CaII infrared
triplet. These lines are much easier to model and have been
successfully inverted in the past few years (e.g., \citeNP{SNTBRC00b};
\citeNP{SNTBRC01}; \citeNP{SN05c}) and thus our results open a new window
to the observational study of EBs.

We find that the EBs analyzed in this work are associated with supersonic (or
nearly so) downflows in the upper photosphere. No upflows are observed in our
data. \citeN{SN05d} determined the vector current densities in the large
sunspot of Fig~\ref{fig:maps} and found that the east penumbra is permeated
by strong currents in the upper photosphere, with a magnitude of
$\sim$3$\times$10$^{5}$~A~km$^{-2}$. The fact that these EBs occur in a
region with strong currents suggests that EBs may be associated with magnetic
reconnection.

In order to understand the nature of EBs and the physical processes that
produce them, it would be extremely valuable to have time-series observations
of photospheric and chromospheric lines, ideally with full Stokes
polarimetry. 

\section{Acknowledgments}

The authors wish to acknowledge the enthusiastic support from the NSO staff
at the Sacramento Peak observatory, especially D. Gilliam, M. Bradford and J.
Elrod. Thanks are also due to S. Hegwer, S. Gregory, R. Dunbar, T. Spence,
S. Fletcher, C. Berst and W. Jones.

%\bibliographystyle{../bib/klunamed}
%\bibliography{../bib/aamnem99,../bib/articulos}

\end{article}

\end{document}